\documentclass[copyright]{eptcs}
\providecommand{\event}{DCFS 2010}
\usepackage[utf8]{inputenc}
\usepackage{xspace}
\usepackage{amsmath}
\usepackage{amssymb}
\usepackage{theorem}
\usepackage{listings}
\usepackage{multirow}
\usepackage{enumerate}
\usepackage{vaucanson-g}
\usepackage{bar}
\usepackage{breakurl}
\newcommand{\DFA}{\textsf{DFA}\xspace}
\newcommand{\DFAs}{\textsf{DFAs}\xspace}
\newcommand{\NFA}{\textsf{NFA}\xspace}
\newcommand{\NFAs}{\textsf{NFAs}\xspace}
\newcommand{\ICDFA}{\textsf{ICDFA}\xspace}
\newcommand{\ICDFAs}{\textsf{ICDFAs}\xspace}
\newcommand{\EFA}{\textsf{EFA}\xspace}
\newcommand{\EFAs}{\textsf{EFAs}\xspace}
\newcommand{\RE}{\mathsf{R}}
\newcommand{\lang}[1]{\mathcal{L}(#1)}
\newcommand{\re}{\textsf{r.e.}\xspace}
\newcommand{\res}{\textsf{r.e.}'s\xspace}

\newcommand{\ie}{\textit{i.e.}\xspace}

\newcommand{\gpath}[2]{\overline{#1\cdots#2}}
\newcommand{\gpatht}[3]{\overline{#1\cdots#2\cdots#3}}

\newcommand{\spg}{{\textsf{SP}}}

\newcommand{\SEA}{\textsf{SEA}\xspace}
\newcommand{\SEAwn}{\textsf{SEAwn}\xspace}
\newcommand{\DMwn}{\textsf{DMwn}\xspace}
\newcommand{\MDMwn}{\textsf{MDMwn}\xspace}
\newcommand{\DM}{\textsf{DM}\xspace}
\newcommand{\HW}{\textsf{HW}\xspace}
\newcommand{\Swn}{\textsf{Swn}\xspace}
\newcommand{\Sn}{\textsf{S}\xspace}
\newcommand{\CS}{\textsf{CS}\xspace}
\newcommand{\MCS}{\textsf{MCS}\xspace}
\newcommand{\CD}{\textsf{CD}\xspace}
\newcommand{\MCD}{\textsf{MCD}\xspace}

\newcommand{\fado}{\textsf{FAdo}\xspace}

\newcommand{\ee}[2]{$#1\times10^{#2}$}
\newcommand{\e}[1]{$#1$}

\title{State Elimination Ordering Strategies: Some Experimental Results}
\author{
Nelma Moreira  \quad\qquad Davide Nabais \quad\qquad Rog\'erio Reis \\
\email{\! nam@ncc.up.pt  \quad\quad\qquad\, dnabais@ncc.up.pt  \quad\qquad rvr@ncc.up.pt} 
\institute{DCC-FC \ \& LIACC,  Universidade do Porto}
\institute{R. do Campo Alegre 1021/1055, 4169-007 Porto, Portugal}}

\def\titlerunning{State Elimination Ordering Strategies} 
\def\authorrunning{N. Moreira and D. Nabais and R. Reis}

\begin{document}

\maketitle

\begin{abstract}
  Recently, the problem of obtaining a short regular expression
  equivalent to a given finite automaton has been intensively
  investigated. Algorithms for converting finite automata to regular
  expressions have an exponential blow-up in the worst-case. To
  overcome this, simple heuristic methods have been proposed.  In
  this paper we analyse some of the heuristics presented in the
  literature and propose new ones.  We also present some experimental
  comparative results based on uniform random generated deterministic
  finite automata.
\end{abstract}

\section{Introduction}
\label{sec:intro}
Recently, the problem of obtaining a short regular expression
equivalent to a given finite automaton has been intensively
investigated. An extensive survey was presented by Ellul \emph{et
  al.}~\cite{ellul05:_regul_expres}, and more recently by Gruber and
Holzer~\cite{gruber08:_provab_short_regul_expres_from}.  It is well
known that the problem of obtaining a minimal regular expression is
PSPACE-complete and NP-complete for acyclic
automata~\cite{jiang93:_minim_nfa}. It is also inefficient to
approximate a minimal regular
expression~\cite{gramlich07:_minim_nfas_and_regul_expres}, unless
P=PSPACE.  Classic algorithms for converting finite automata to
regular expressions can produce regular expressions of size ${\cal
  O}(nk4^n)$ in the worst case, where $n$ is the number of states and
$k$ the alphabet size of the correspondent automaton. Several
exponential lower bounds are provided in the
literature~\cite{ellul05:_regul_expres,gruber08:_finit_autom_digrap_connec_and}
showing that the exponential blow-up is unavoidable.  For specific
classes of automata, better upper bounds can be
found~\cite{ellul05:_regul_expres,gulan08:_local_elimin_strat_in_autom_short,sakarovitch05:_languag_expres_small_autom,moreira09:_series_paral_autom_and_short_regul_expres}.
In particular, Gruber and
Holzer~\cite{gruber08:_provab_short_regul_expres_from} presented an
algorithm that converts an $n$-state deterministic finite automaton
(\DFA{}) over a binary alphabet into a regular expression of size at
most ${\cal O}(1.742^n)$.  In general, to obtain shorter regular
expressions it is essential the order in which the automaton's
states are considered in the conversion.  To tackle the problem of
obtaining an optimal ordering in a feasible manner, heuristic methods
have been
proposed~\cite{delgado04:_approx_to_small_regul_expres,han07:_obtain_short_regul_expres_finit_state_autom,ahn09:_implem_of_state_elimin_using_heuris}.

In this paper we analyse some of the heuristics presented in the
literature and propose new ones. To test their performance, some
experimental results were carried out using statistically significant
samples obtained with an uniform random generator. The paper is
organized as follows. In the next section some basic notions are
reviewed. Section 3 summarizes the conversions from finite automata to
regular expressions, and in particular the state elimination
method. Section 4 describes some elimination ordering strategies and
two new ones are proposed. In Section 5 experimental results are
analysed and Section 6 concludes.

\section{Preliminaries}
\label{sec:pre}

We recall some basic notions of digraphs, finite automata and regular
expressions. For more details we refer the reader to standard
books~\cite{hopcroft00:_introd_autom_theor_languag_comput,sakarovitch09:_elemen_of_autom_theor,harary69:_graph_theor}.

A digraph $D=(V,E)$ consists of a finite set $V$ of vertices and a set
$E$ of ordered pairs of vertices, called $arcs$.  If $(u,v)$ in $E$,
$u$ is \emph{adjacent to} (or incident to) $v$ and $v$ is
\emph{adjacent from} $u$.  For each vertex $v$, the \emph{indegree} of
$v$ is the number $n_i$ of vertices adjacent to it and the
\emph{outdegree} of $v$ is the number $n_o$ of vertices adjacent from
it, and we write $v(n_i;n_o)$. An arc $(u,v)$ can be denoted by $uv$.
A \emph{path} between $v_0$ and $v_n$ is a sequence
$v_0v_1,v_1v_2,\ldots,v_{n-1}v_n$ of arcs, and is denoted by
$\gpath{v_0}{v_n}$, or $\gpatht{v_0}{v_k}{v_n}$, for $1\leq k< n$. A
path is \emph{simple} if all the vertices in it are distinct. The
length of a path is the number of arcs in the path. A path is a
\emph{cycle} if $v_0=v_n$ and $n\geq 1$. A digraph that has no cycles
is called \emph{acyclic}.

We now review some notions and notation from formal languages and
finite automata. Let $\Sigma$ be a finite alphabet and
$\Sigma^\star$ be the set of words over $\Sigma$. The empty word is
denoted by $\varepsilon$. A \emph{language} over $\Sigma$ is a
subset of $\Sigma^\star$.  A \emph{regular expression} (\re)
$\alpha$ over $\Sigma$ represents a regular language
$\lang{\alpha}\subseteq \Sigma^\star$ and is inductively defined
by: $\emptyset$ is a \re and $\lang{\emptyset}=\emptyset$;
$\varepsilon$ is a \re and $\lang{\varepsilon}=\{\varepsilon\}$; $a\in
\Sigma$ is a \re and $\lang{a}=\{a\}$; if $\alpha_1$
and $\alpha_2$ are \re, $(\alpha_1 + \alpha_2)$, $(\alpha_1 \alpha_2)$,
and $(\alpha_1)^\star$ are \re, respectively with $\lang{(\alpha_1 +
  \alpha_2)}=\lang{\alpha_1}\cup \lang{\alpha_2}$,
$\lang{(\alpha_1\alpha_2)}=\lang{\alpha_1}\lang{\alpha_2}$, and
$\lang{(\alpha_1)^\star}=\lang{\alpha_1}^\star$.  The \emph{alphabetic
  size} of an \re $\alpha$ is the number of alphabetic symbols of $\alpha$ and is
denoted by $|\alpha|_\Sigma$.  Let $\RE$ be the set of regular
expressions over $\Sigma$.  Two regular expressions $\alpha$ and
$\beta$ are \emph{equivalent} if $\lang{\alpha}=\lang{\beta}$, and we
write $\alpha=\beta$. With this interpretation, the algebraic
structure $(\RE,+,\cdot,\emptyset,\varepsilon)$ constitutes an
idempotent semiring, and with the unary operator $\star$, a Kleene
algebra.  Using these algebraic properties as (simplification) rewrite
rules, it is possible to decide if two regular expressions are
equivalent, but no algorithm is known to minimize a given
regular expression (except a \texttt{brute-force} one).  

A {\em non-deterministic finite automaton} (\NFA)
$\mathcal{A}$ is a quintuple $(Q,\Sigma,\delta,q_0,F)$ where $Q$ is
a finite set of states, $\Sigma$ is the alphabet, $\delta \subseteq
Q \times \Sigma \times Q$ is the transition relation, $q_0$ the
initial state and $F \subseteq Q$ is the set of final states. For
$q\in Q$ and $a\in \Sigma$, we denote the set
$\{p\in Q\mid(q,a,p)\in \delta\}$ by $\delta(q,a)$, and we can
extend this notation to $w\in \Sigma^\star$, and to $R\subseteq Q$.
The \emph{language} recognized by $\mathcal{A}$ is
$\lang{\mathcal{A}}=\{w\in \Sigma^\star\mid \delta(q_0,w)\cap
F\not= \emptyset\}$.  An \NFA is \emph{deterministic} (\DFA) if for
each pair $(q,a) \in Q \times \Sigma$, $|\delta(q,a)|\leq 1$. A \DFA
is complete if $\delta$ is a total function.  An \NFA is
\emph{initially-connected} if for each state $q\in Q$ there exists a
word $w\in \Sigma^\star$ such that $q\in \delta(q_0,w)$. A complete
initially-connected \DFA is denoted by \ICDFA.  An \NFA is \emph{trim}
if it is initially-connected and if every state is \emph{useful}, \ie,
for all $q\in Q$ there exist a word $w\in \Sigma^\star$ such that
$F\cap\delta(q,w)\not=\emptyset$.  The \emph{underlying digraph} of an
\NFA ${\cal A} = (Q, \Sigma, \delta,q_0,F)$ is the digraph $D=(Q,E)$ such that
$E=\{ (q,q')\mid q,q'\in Q \text{ and } \exists a\in
\Sigma\cup\{\varepsilon\} \text{ such that } (q,a,q')\in \delta\}$. Note
that even if there can be more than one symbol of $\Sigma$ between two
states $q$ and $q'$, only one arc exists in the underlying digraph. 

For the conversion from \NFAs to \res extended finite automata are
considered. An \emph{extended finite automaton} (\EFA) ${\cal A}$ is a
quintuple $(Q,\Sigma,\delta,q_0,F)$, where $Q$, $\Sigma$, $q_0$ and
$F$ are as before, and $\delta:Q\times Q\rightarrow \RE$.  We assume
that $\delta(q,q')=\emptyset$, if the transition from $q$ to $q'$ is
not defined. Any \NFA can be easily transformed into an equivalent \EFA, with the same
underlying digraph: for each pair of states $(q,q')$ one needs to
construct a regular expression $a_1+\cdots+ a_n$ such that
$(q,a_i,q')\in \delta$, $a_i\in \Sigma\cup \{\varepsilon\},\, 1\leq i\leq
n$. This transformation corresponds to eliminate \emph{parallel}
transitions. Whenever appropriated we will use the same terminology
both for digraphs and for automata.

\section{From Finite Automata to Regular Expressions}
\label{sec:fa2re}

Kleene's theorem~\cite{kleene56:_repres} establishing the equivalence
between languages accepted by finite automata and represented by
regular expressions provided proof that a language accepted by an \NFA
can be represented by a \re. McNaughton and
Yamada~\cite{mcnaughton60:_regul_expres_and_state_graph_for_autom}
presented a recursive algorithm that calculates a \re from an \NFA
based on the computation of the transitive closure of the underlying
digraph. Brzozowski and
McCluskey~\cite{brzozowski63:_signal_flow_graph_techn_for} introduced
a method now known as \emph{state elimination algorithm} (\SEA) that
considers \EFAs and leads, in general, to simpler computations and
shorter \res. A third method exists based on solving a system of
linear equations akin a Gaussian elimination
process~\cite{arden60:_delay_logic_and_finit_state_machin,kozen94:_kleen}.
This last approach is interesting as linear algebra or optimization
techniques can be adapted in order to provide new methods to obtain
\res.
Sakarovitch~\cite{sakarovitch05:_languag_expres_small_autom,sakarovitch09:_elemen_of_autom_theor}
studied the relationship between the three methods and in particular
showed that given an order in the set of states $Q$ the regular
expressions obtained by two different methods can be reduced to each
other by the application of a specific subset of algebraic properties.
 
Most improvements and heuristic methods are based on the state
elimination method and try to identify state orderings that lead to
shorter \res.

\subsection{State Elimination Method Revisited}
\label{sec:sea}
The state elimination algorithm takes as input an \EFA and produces an
equivalent \re. In each step, a non-initial and non-final state of the
\EFA is eliminated (deleted) and the transitions are changed in such
way that the new and the older \EFAs are equivalent.  Usually it is
assumed that the input \EFA is trim and \emph{normalized}, \ie, the
initial state has no incoming transitions, there is only a final
state and that state has no outgoing transitions. Every \EFA (or \NFA) can be
transformed into an equivalent normalized \EFA. Formally, let
${\cal A}=(Q,\Sigma,\delta,q_o,F)$ be an \EFA, then:

\begin{description}
\item[Normalization:] \ 

\begin{enumerate}[(NI)]
\item  \label{nor:i}If there is $q\in Q$ such that
  $\delta(q,q_0)\not=\emptyset$, then add a new state $i$ to $Q$,
  define $\delta(i,q_0)=\varepsilon$, and set $i$ as the new initial
  state.
\item \label{nor:f}If $|F|> 1$ or exists $q\in F$ and $q' \in Q$ such that
  $\delta(q,q')\not=\emptyset$, then add a new state $f$ to $Q$ and a
  transition $\delta(q,f)=\varepsilon$, for all $q\in F$. The set of
  final states becomes $\{f\}$.
\end{enumerate}
\end{description}

Without lost of generality, let $A'=(Q',\Sigma,\delta',i,f)$ denote
the new normalized \EFA. Let $\alpha_{qq'}$ denote the regular
expression $\delta(q,q')$. Normalization is preserved when the below
\emph{state elimination} process is performed.
\begin{description}
\item[State Elimination:]\ \label{sea}

  \begin{enumerate}[(EI)]
\item  \label{se} If $Q=\{i,f\}$, then the resulting regular
  expression is $\alpha_{if}$, and the
  algorithm terminates. Otherwise continue to step~(E\ref{el}).
\item\label{el} \textbf{Choose} $q\in Q\setminus\{i,f\}$.
  Eliminate $q$ from $A'$, considering $Q'\setminus\{q\}$ the new set of
  states, and for each $q_1,q_2\in Q'\setminus\{q\}$,
  $$\delta'(q_1,q_2)=\alpha_{q_1q_2}+\alpha_{q_1q}\alpha_{qq}^\star \alpha_{qq_2},$$
  Continue to step~(E\ref{se}).
  \end{enumerate}
\end{description}

Hopcroft \emph{et
  al.}~\cite{hopcroft00:_introd_autom_theor_languag_comput}
presented a slight variation of the above algorithm that omits the
normalization step.  Considering that there is only one final state,
state elimination ends with one of the following \EFAs (where some \res
can be $\emptyset$):
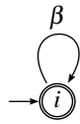
\begin{center}
\begin{tabular}[c]{ccccccc}
\SmallPicture{\VCDraw{%
\begin{VCPicture}{(-3,-1)(3,3)}
\FinalStateVar[i]{(0,0)}{1}
\Initial{1}
\LoopN[0.5]{1}{\beta}
\end{VCPicture}
}}
 &&&&&&
\SmallPicture\VCDraw{%
\begin{VCPicture}{(-3,-2)(3,3)}
\State[i]{(-1,0)}{1}
\Initial{1}
\FinalState[f]{(2,0)}{3}
\LoopN[0.5]{1}{\beta_1}
\ArcL	[0.5]{1}{3}{\alpha_1}
\ArcL	[0.5]{3}{1}{\alpha_2}
\LoopN[0.5]{3}{\beta_2}
\end{VCPicture}
}
\\
 Initial state is final. &&&&&& There are two different states. 
 \end{tabular}
  \end{center}
  \noindent In the left case, the final regular expression is
  $\beta^\star$ and in the right case, the final regular expression
  can be
  $\beta_1^\star\alpha_1(\beta_2+\alpha_2\beta_1^\star\alpha_1)^\star$
  or any shorter \re if some of the transitions are labelled by
  $\emptyset$.  When $|F|>1$ the normalization step~(N\ref{nor:f})
  should be considered. We refer, by abuse of language, to this algorithm
  as the \SEA \emph{without normalization} (\SEAwn). It has the
  advantage of avoiding unnecessary $\varepsilon$ transitions, and, as
  we will see in Section~\ref{sec:nn}, it exhibits a better
  performance for  the elimination strategies.

\section{State Elimination Orderings}
\label{sec:seo}
The importance of the order in which the states are considered in
the conversion, was noticed by the authors of the early algorithms.
McNaugthon and Yamada suggested that states with higher in- and
outdegrees should be considered at the end. Brzozowski and McCluskey
proposed to eliminate first the states $q\in Q$ such that $q(1;1)$,
\ie, $q$ connects two other states in \emph{series}:
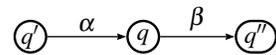
\begin{center}
\SmallPicture{\VCDraw{%
\begin{VCPicture}{(-3,-1)(3,1)}
\State[q']{(-3,0)}{1}
\State[q]{(0,0)}{2}
\StateVar[q'']{(3,0)}{3}
\EdgeL[0.5]{1}{2}{\alpha}
\EdgeL[0.5]{2}{3}{\beta}
\end{VCPicture}
}}

\end{center}
Acyclic \NFAs for which in each step of the state elimination process
there is a state satisfying these conditions were studied by Moreira and
Reis~\cite{moreira09:_series_paral_autom_and_short_regul_expres} and
called \spg-automata. For this class it is possible to obtain a linear
size \re in ${\cal O}(n^2\log n)$ time. If an acyclic \NFA is not \spg, it
must be reduced by series-parallel elimination to one that contains a
subgraph of the form:
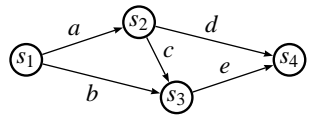
\begin{center}
  \SmallPicture{\VCDraw{
      \begin{VCPicture}{(0,-1)(7,3)}
        \State[s_1]{(0,1)}{A} \State[s_2]{(3,2)}{B}
        \State[s_3]{(4,0)}{C} \State[s_4]{(7,1)}{D} 
        \EdgeL{A}{B}{{a}} \EdgeR{A}{C}{{b}}
        \EdgeL{B}{C}{{c}} \EdgeL{B}{D}{{d}}
        \EdgeL{C}{D}{{e}}
      \end{VCPicture}}}
\end{center}
And, in general, it is not easy to see which elimination ordering should
be considered.

The \spg-automata strategy was extended by Gulan and
Fernau~\cite{gulan08:_local_elimin_strat_in_autom_short} for a
specific case of cyclic \NFAs.  \spg-automata belong to the class of
graphs which excludes a complete graph as a minor. For this class,
Ellul and \emph{et al.} proved that there are \res which size is less
than $e^{{\cal O}(\sqrt{n})}$. Gruber and Holzer extended this work
to \DFAs, providing an algorithm with a guaranteed performance of
${\cal O}(1.742^n)$ for binary alphabets.

\subsection{Delgado and Morais Heuristics}
\label{sec:dm}
In each step of the state elimination process, given
$q(m;l)$, the contribution of this state  for the size of the final
regular expression can be measured by
\begin{equation}
  \label{eq:weight}
  W(q)=(l-1)\sum_{i=1}^{m}|\alpha_{q_iq}|+(m-1)\sum_{j=1}^{l}|\alpha_{qq_j}|+(ml
  -1)|\alpha_{qq}|.
\end{equation}

Delgado and Morais~\cite{delgado04:_approx_to_small_regul_expres}
proposed a strategy (\DM) that in each step eliminates a state $q$
with the lowest \emph{weight} $W(q)$. Although this heuristic  is  quite simple
 and runs in ${\cal O}(n^2)$, the experimental results
provides evidence that it has  very good performance. Recently, Gruber~{\em et
  al.}~\cite{gruber09:_short_regul_expres_from_finit_autom} presented
more experimental results which showed statistical significance and
were based on
uniform random generated \ICDFAs, where this heuristic almost always
outperforms several others. Our results corroborate this good
performance. In particular, when applied to an \spg-automaton, this
heuristics always selects a state $q$ such that $q(1;1)$, producing a
linear size \re.

\subsection{Han and Wood Heuristics}
\label{sec:han}
Han and Wood~\cite{han07:_obtain_short_regul_expres_finit_state_autom}
introduced the notion of \textit{bridge state} which leads to a
decomposition of the \EFA, therefore of the elimination process.  That
notion was redefined by Ahn and
Han~\cite{ahn09:_implem_of_state_elimin_using_heuris}, as follows: a
state $q$ is a \textit{bridge state} if it satisfies the following
conditions:
 \begin{enumerate}[(BI)]
 \item   \label{b1} $q$ is neither initial nor final;
 \item \label{b2} For any $f\in F$, each path $\gpath{i}{f}$ must pass
   through $q$, \ie, must be of the form $\gpatht{i}{q}{f}$, where $i$
   is the initial state;
\item \label{b3} $q$ does not participate in any cycle except for a loop.
\end{enumerate}
Note that bridge states correspond to the usual notion of \emph{cut
  points}, with the extra constraint~(B\ref{b3}). Bridge states can be
found in linear time, and it was proved that in an optimal elimination
ordering the bridge states must be the last ones.  This is easy to see
because the automaton can be \emph{decomposed} into two subautomata
${\cal A}_1$ and ${\cal A}_2$, such that a bridge state $q$ corresponds to the final
state of ${\cal A}_1$ and the initial state of ${\cal A}_2$:
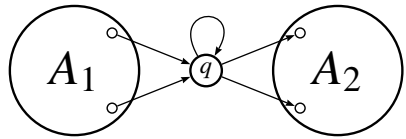
\begin{center}
\SmallPicture{
\VCDraw{%
\begin{VCPicture}{(-5,-3)(9,3)}
\FixStateDiameter{3.5cm}\ChgStateLabelScale{2}
\State[A_1]{(-1,0)}{A}
\State[A_2]{(6,0)}{B}
\MediumState\RstStateLabelScale
\State[q]{(2.5,0)}{q}
\VSState{(0,-1)}{p1}\VSState{(0,1)}{p2}
\EdgeL{p1}{q}{}\EdgeL{p2}{q}{}
\VSState{(5,-1)}{p3}\VSState{(5,1)}{p4}
\EdgeL{q}{p4}{}\EdgeL{q}{p3}{}
\LoopN{q}{}
\end{VCPicture}
}}
\end{center}
Ahn and Han present some empirical results of this strategy (that we
designed by \HW) combined
with the one based on state weights (\DM) and also with one that
performs a \emph{parallel} decomposition of the \EFA. Although the
dataset used is random generated, it is not uniform nor statistically
significant.

\subsection{\SEA Without Normalization}
\label{sec:nn}

Consider the following simple \DFA:
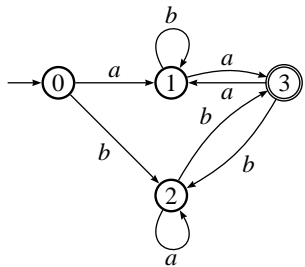
\begin{center}
\SmallPicture{\VCDraw{%
\begin{VCPicture}{(-4,-4)(4,4)}
\State[1]{(0,2)}{1}
\FinalState[3]{(3,2)}{3}
\State[0]{(-3,2)}{7}
\Initial[w]{7}
\State[2]{(0,-1)}{9}
\LoopN[0.5]{1}{b}
\ArcL	[0.5]{1}{3}{a}
\EdgeL[0.5]{3}{1}{a}
\EdgeL[0.5]{7}{1}{a}
\LoopS[0.5]{9}{a}
\ArcL	[0.5]{3}{9}{b}
\ArcL	[0.5]{9}{3}{b}
\EdgeR[0.5]{7}{9}{b}
\end{VCPicture}
}}

\end{center}
Applying the \SEA with normalization to this \DFA and using the \DM
strategy, the first state to be eliminate corresponds to the initial
state (\ie it is the one with small weight). This will lead to a \re
with the highest alphabetic size ($29$), within all that can be
obtained by state elimination. The elimination ordering is $0$,
$3$, $1$, $2$.

 On the other hand, if we consider a
\SEA with the Hopcroft {\em et al.} approach (such that the initial
state is only considered at the end) applying the \DM strategy will
lead to a \re with the smallest alphabetic size ($12$). Now, the
elimination ordering is $2$, $1$ (as the two other states are fixed). This strategy
corresponds to combine the \DM strategy with one where the initial
state is the last to be eliminated. Our experimental results below show that
this approach (\SEAwn) improves, in general, the strategies we considered.

\subsection{A New  Heuristic: Counting Cycles}
\label{sec:cscd}
Consider, now, the following \DFA
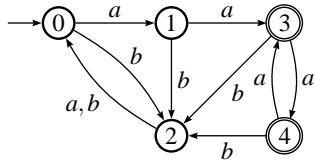
\begin{center}
\SmallPicture{\VCDraw{%
\begin{VCPicture}{(-5,-2)(6,4)}
\State[1]{(0,2)}{0}
\FinalState[3]{(3,2)}{1}
\State[0]{(-3,2)}{2}
\Initial[w]{2}
\State[2]{(0,-1)}{3}
\FinalState[4]{(3,-1)}{14}
\EdgeL[0.5]{2}{0}{a}
\ArcL	[0.5]{2}{3}{b}
\ArcL	[0.5]{3}{2}{a,b}
\EdgeL[0.5]{0}{3}{b}
\ArcL	[0.5]{1}{14}{a}
\ArcL	[0.5]{14}{1}{a}
\EdgeL[0.5]{1}{3}{b}
\EdgeL[0.5]{14}{3}{b}
\EdgeL[0.5]{0}{1}{a}
\end{VCPicture}
}}

  \end{center}
  The \DM heuristics produces a \re with alphabetic size $29$ or $26$,
  if either \SEA or \SEAwn is considered. The corresponding
  elimination order are $1$, $4$, $0$, $2$, $3$ and \e{1}, \e{3}, $2$,
  $4$, respectively. For this \DFA the optimal alphabetic size for \re
  obtained by the state elimination method is $16$ (and the worst is
  $126$). Instead of the weight of a state being the weighted summation
  of its in- and out-degrees, one can consider the number of cycles
  that pass through it (multiplicities included). In this particular
  case the obtained \re has size $19$. The number of cycles for each
  state is, by increasing identifier order, 4, 3, 4, 3 and 2,
  respectively.

  Two strategies can be developed to obtain an elimination
  ordering:
\begin{enumerate}[(CI)]
\item\label{cs} statically determine the number of cycles for each
  state $q$, of the original automaton (\CS); this can be achieved in
  ${\cal O}(n^2)$.
\item\label{cd} dynamically determine those values after each
  elimination step (\CD); this can be achieved in ${\cal O}(n^3)$.
\end{enumerate}
In the second case, (C\ref{cd}), instead of the multiplicities, the
alphabetic size of each transition label is considered.

\section{Experimental Results}
\label{sec:results}

Each of the state elimination algorithms described before was implemented in
Python within the \fado
system~\cite{moreira05:_inter_manip_regul_objec_fado,almeida09:_fado_guitar,fado}.
The experiments were undertaken with samples of $10,000$ uniform
random generated
\ICDFAs~\cite{almeida07:_enumer_gener_strin_autom_repres} with a fixed
number of states ($n$) and alphabet size ($k$).  The sample size
ensures the statistical significance with a 95\% confidence level
within a 1\% error margin. Most of the tests were performed for
automata with $n\in \{10,20,50\}$ states and $k\in
\{2,3,5,10,26,100\}$ symbols. Each generated automaton is represented
by a \emph{canonical} string. Assuming an ordering on the alphabet,
the states are numbered from $0$ to $n-1$,
$0$ being the initial state.  The string representation is a list of
states reached from each state by increasing order of symbols and of
state numbering, beginning with
the initial state. For example, the string for the \DFA of
Section~\ref{sec:nn}, considering $a<b$, is $12312312$.

Experiments were carried out considering the  following goals:
\begin{itemize}
\item to determine the density of occurrence of bridge states in
  (complete) \DFAs.
\item to test the performance of \SEAwn, \ie\ the state elimination method
  without normalization, independently of other elimination ordering
  strategies;
\item to test the performance of the strategies based on counting the
  number of cycles.
\end{itemize}

\subsection{Bridge States Density}
\label{sec:ocuts}

The performance of the strategy \HW proposed by Han and Wood, and
described in Section~\ref{sec:han}, heavily depends on the existence
of bridge states in a finite automaton. We estimated the occurrence of
these states in \ICDFAs, and their average position in the \ICDFA
canonical string. In the string representation, an early position
corresponds to a closer proximity to the initial state. Thus this
index measures the state distance from the initial state and gives
information about the number of states of each subautomaton in which
the \ICDFA can be decomposed.  In the following table, and for each
sample, \emph{tot} is the total number of bridge states, \emph{num} is
the number of \ICDFAs with at least a bridge state and \emph{pos} is
their average position in the \ICDFA canonical string. The table
values suggest that bridge states are very rare and a bridge state is
usually the initial state or adjacent from it. Note that for larger
alphabets ($k\geq 10$) no bridge states, at all, were found.

\medskip

{\small
\begin{center}
  \begin{tabular}{|c||r|r|r||r|r|r||r|r|r||r|r|r|}
    \hline
    &\multicolumn{3}{c||}{$k=2$}&\multicolumn{3}{c||}{$k=3$}&\multicolumn{3}{c||}{$k=5$}&\multicolumn{3}{c|}{$k=10$}\\\hline
    &tot&num&pos&tot&num&pos&tot&num&pos&tot&num&pos\\\hline
    $n=10$&$3252$&$2327$&$0.824$&$829$&$707$&$0.458$&$88$&$82$&$0.193$&$0$&$0$&N/A\\
    $n=20$&$3506$&$2375$&$1.224$&$757$&$634$&$0.486$&$73$&$71$&$0.123$&$0$&$0$&N/A\\
    $n=50$&$3499$&$2411$&$1.375$&$758$&$649$&$0.451$&$69$&$63$&$0.115$&$0$&$0$&N/A\\\hline
  \end{tabular}
\end{center}
}
\medskip

\subsection{\SEAwn Performance}
\label{sec:norm-cons}

To test the performance of the \SEAwn method, several elimination
ordering strategies were considered. A trivial order is the one in
which the states occur in the \ICDFA canonical string. This ordering
produces very bad results (even compared with a random one) but here
we wanted to test the effect of the prior automata normalization. The
correspondent algorithms are \Sn and \Swn, respectively. We also
considered the \DM strategy with the \SEAwn method (\DMwn). For each
pair of algorithms, the ratio between the average \re alphabetic sizes
was computed.  The following bar charts summarize some of the results.
The \Swn method (without normalization) always outperforms the \Sn
(with normalization). Because the \re sizes are huge some ratios are
very small. For example, a ratio of $0.08$, for $n=50$ and $k=10$,
corresponds to the diminishing of two orders of magnitude (from
$10^{27}$ to $10^{25}$). The \DMwn method can achieve an improvement
of $15\%$ over the \DM one. 

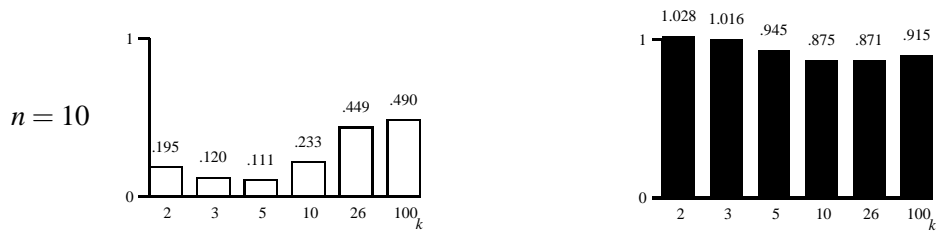
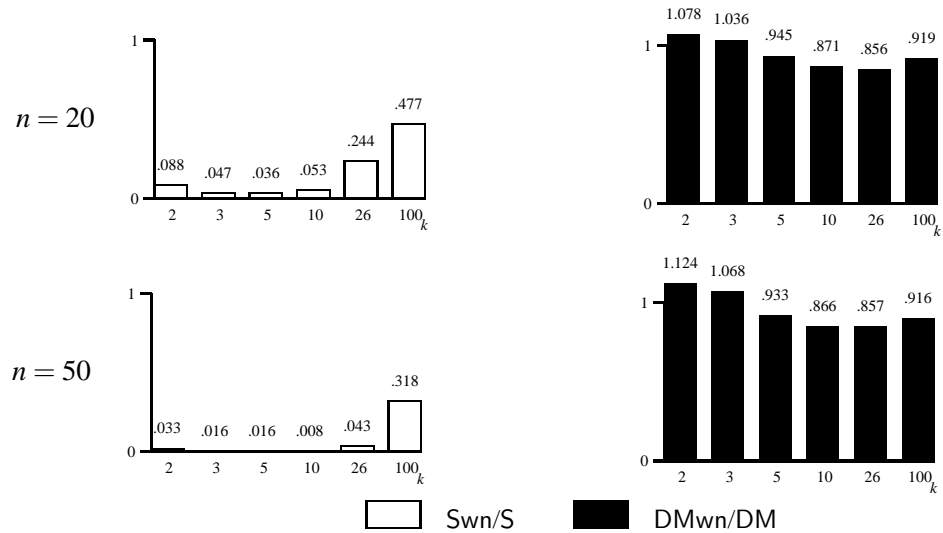
\medskip

\begin{center}
$n=10$
  \begin{tabular}{cc}
    \begin{minipage}{.4\linewidth}
      \begin{barenv}
        \setwidth{12}
        \sethspace{.5}
        \setstretch{60}
        \setstyle{\tiny}
        \setyaxis{0}{1.0}{1}
        \setnumberpos{up}
        \setprecision{2}
        \setxname{\tiny $k$}
        \bar{.195}{1}[{\tiny{} 2}]
        \bar{.120}{1}[{\tiny{} 3}]
        \bar{.111}{1}[{\tiny{} 5}]
        \bar{.233}{1}[{\tiny{} 10}]
        \bar{.449}{1}[{\tiny{} 26}]
        \bar{.490}{1}[{\tiny{} 100}]
      \end{barenv}
    \end{minipage}
    &
    \begin{minipage}{.4\linewidth}
    \begin{barenv}
      \setwidth{12}
      \sethspace{.5}
      \setstretch{60}
      \setstyle{\tiny}
      \setyaxis{0}{1.0}{1}
      \setnumberpos{up}
      \setprecision{2}
      \setxname{\tiny $k$}
      \bar{1.028}{8}[{\tiny{} 2}]
      \bar{1.016}{8}[{\tiny{} 3}]
      \bar{.945}{8}[{\tiny{} 5}]
      \bar{.875}{8}[{\tiny{} 10}]
      \bar{.871}{8}[{\tiny{} 26}]
      \bar{.915}{8}[{\tiny{} 100}]
    \end{barenv}
    \end{minipage}
  \end{tabular}
\end{center}
\vspace{.5cm}
\begin{center}
  $n=20$
  \begin{tabular}{cc}
    \begin{minipage}{.4\linewidth}
      \begin{barenv}
        \setwidth{12}
        \sethspace{.5}
        \setstretch{60}
        \setstyle{\tiny}
        \setyaxis{0}{1.0}{1}
        \setnumberpos{up}
        \setprecision{2}
        \setxname{\tiny $k$}
        \bar{.088}{1}[{\tiny{} 2}]
        \bar{.047}{1}[{\tiny{} 3}]
        \bar{.036}{1}[{\tiny{} 5}]
        \bar{.053}{1}[{\tiny{} 10}]
        \bar{.244}{1}[{\tiny{} 26}]
        \bar{.477}{1}[{\tiny{} 100}]
      \end{barenv}
    \end{minipage}
    &
    \begin{minipage}{.4\linewidth}
    \begin{barenv}
      \setwidth{12}
      \sethspace{.5}
      \setstretch{60}
      \setstyle{\tiny}
      \setyaxis{0}{1.0}{1}
      \setnumberpos{up}
      \setprecision{2}
      \setxname{\tiny $k$}
      \bar{1.078}{8}[{\tiny{} 2}]
      \bar{1.036}{8}[{\tiny{} 3}]
      \bar{.945}{8}[{\tiny{} 5}]
      \bar{.871}{8}[{\tiny{} 10}]
      \bar{.856}{8}[{\tiny{} 26}]
      \bar{.919}{8}[{\tiny{} 100}]
    \end{barenv}
    \end{minipage}
  \end{tabular}
\end{center}
\vspace{.5cm}
\begin{center}
  $n=50$
  \begin{tabular}{cc}
    \begin{minipage}{.4\linewidth}
      \begin{barenv}
        \setwidth{12}
        \sethspace{.5}
        \setstretch{60}
        \setstyle{\tiny}
        \setyaxis{0}{1.0}{1}
        \setnumberpos{up}
        \setprecision{2}
        \setxname{\tiny $k$}
        \bar{.033}{1}[{\tiny{} 2}]
        \bar{.016}{1}[{\tiny{} 3}]
        \bar{.016}{1}[{\tiny{} 5}]
        \bar{.008}{1}[{\tiny{} 10}]
        \bar{.043}{1}[{\tiny{} 26}]
        \bar{.318}{1}[{\tiny{} 100}]
      \end{barenv}
    \end{minipage}
    &
    \begin{minipage}{.4\linewidth}
    \begin{barenv}
      \setwidth{12}
      \sethspace{.5}
      \setstretch{60}
      \setstyle{\tiny}
      \setyaxis{0}{1.0}{1}
      \setnumberpos{up}
      \setprecision{2}
      \setxname{\tiny $k$}
      \bar{1.124}{8}[{\tiny{} 2}]
      \bar{1.068}{8}[{\tiny{} 3}]
      \bar{.933}{8}[{\tiny{} 5}]
      \bar{.866}{8}[{\tiny{} 10}]
      \bar{.857}{8}[{\tiny{} 26}]
      \bar{.916}{8}[{\tiny{} 100}]
    \end{barenv}
    \end{minipage}
  \end{tabular}
\vspace{.5cm}

{\small \legend1{\Swn/\Sn}\quad\quad{}\legend8{\DMwn/\DM}}
\end{center}

\subsection{Cycle Heuristic Performance}
\label{sec:cycle-heur-perf}
The two heuristics presented in Section~\ref{sec:cscd}, \CS and \CD,
were implemented using the \SEAwn method. It was then natural to
compare their performance with \DMwn, the best heuristic so far. The
following table summarizes the results. The third to the fifth columns
have the average \re alphabetic sizes obtained for each of the
mentioned heuristics. The sixth column corresponds to the average of
the minimum value of the three, \emph{the best of the 3}
(\textsf{B3}). The three last columns contain the maximum values
obtained by each of the heuristics. 

\medskip

\begin{small}
  \begin{center}
    \begin{tabular}{|c|c|r|r|r|r|r|r|r|}
      \hline
      $k$&$n$&\DMwn&\CS&\CD&\textsf{B3}&\MDMwn&\MCS&\MCD\\\hline\hline
      $2$&$10$&$149$&$144$&$143$&$135$&$864$&$1014$&$909$\\
      &$20$&$1557$&$1531$&$1617$&$1331$&$12494$&$18235$&$16230$\\
      &$50$&\ee{3.5}{5}&\ee{4.9}{5}&\ee{5.5}{5}&\ee{2.5}{5}&\ee{7.8}{6}&\ee{1.9}{7}&\ee{2.5}{7}\\\hline
      $3$&$10$&\e{633}&\e{617}&\e{628}&\e{564}&\e{4792}&\e{4206}&\e{5095}\\
      &\e{20}&\e{23431}&\e{25817}&\e{27560}&\e{18739}&\e{339595}&\e{365533}&\e{428164}\\
      &\e{50}&\ee{2.5}{8}&\ee{7.6}{8}&\ee{6.5}{8}&\ee{1.6}{8}&\ee{1.0}{10}&\ee{1.6}{11}&\ee{8.9}{10}\\\hline
      \e{5}&\e{10}&\e{4492}&\e{4646}&\e{4713}&\e{3942}&\e{32780}&\e{34044}&\e{35508}\\
      &\e{20}&\ee{1.0}{6}&\ee{1.5}{6}&\ee{1.4}{6}&\ee{8.2}{5}&\ee{1.2}{7}&\ee{2.8}{7}&\ee{2.7}{7}\\
      &\e{50}&\ee{5.5}{12}&\ee{3.5}{13}&\ee{2.0}{13}&\ee{3.2}{12}&\ee{4.4}{14}&\ee{5.3}{15}&\ee{3.1}{15}\\\hline
      \e{10}&\e{10}&\e{52943}&\e{59921}&\e{57138}&\e{47564}&\e{232338}&\e{430391}&\e{262446}\\
      &\e{20}&\ee{1.8}{8}&\ee{3.1}{8}&\ee{2.7}{8}&\ee{1.4}{8}&\ee{1.7}{9}&\ee{9.9}{9}&\ee{3.2}{9}\\\hline
      \e{26}&\e{10}&\ee{6.0}{5}&\ee{7.1}{5}&\ee{6.5}{5}&\ee{5.8}{5}&\ee{1.1}{6}&\ee{1.7}{6}&\ee{1.5}{6}\\
      &\e{20}&\ee{3.3}{10}&\ee{5.7}{10}&\ee{4.4}{10}&\ee{2.9}{10}&\ee{1.3}{11}&\ee{3.8}{11}&\ee{1.8}{11}\\\hline
      \e{100}&\e{10}&\ee{4.1}{6}&\ee{4.2}{6}&\ee{4.1}{6}&\ee{4.1}{6}&\ee{5.3}{6}&\ee{5.6}{6}&\ee{5.5}{6}\\
      &\e{20}&\ee{1.5}{12}&\ee{1.7}{12}&\ee{1.6}{12}&\ee{1.4}{12}&\ee{2.1}{12}&\ee{2.9}{12}&\ee{2.7}{12}\\\hline
    \end{tabular}
  \end{center}
\end{small}

\medskip

On average, the heuristics \DMwn outperforms the other two, although
not always. However, the performance of the cycle heuristics are of
the same order of magnitude.  The comparison between \CS and \CD is
hard to interpret. The overhead of reevaluate the cycle weights after
each step seems not worthwhile. This suggest that the \CS strategy is
a good choice, even compared with \DMwn, as the weights are computed
only once. The most important result is that considering the three
heuristics a better value is always obtained (\textsf{B3}). This means that
when \DMwn produces a bad value one of the other two produces a better
value, and vice versa. This is surprising, and deserves future research.

\section{Conclusions}
\label{sec:conc}
Several state elimination ordering strategies were analysed and new
ones were proposed.  Experimental results were conducted with
statistical accurate samples of uniform random generated
deterministic finite automata. In this context the following
conclusions can be drawn:
\begin{itemize}
\item a general improvement in all strategies is obtained if the \SEA
  \emph{without normalization} is considered;

\item bridge states are very rare;

\item the \HW strategy clearly clash with the new strategies based on
  the number of cycles count (\CS and \CD), because bridge states are
  cycle free; but, as we saw, their rarity makes this contradiction
  unimportant;

\item the new proposed strategies (\CS and \CD) are comparable with
  the \DM heuristic; however these new heuristics only outperform, on
  average, the \DM heuristic for automata with small alphabets and
  small number of states;

\item if one takes as strategy, for each automaton, the best result from
  these three heuristics (\DM, \CS and \CD)  a gain of 25\% is
  obtained, with the same worst case complexity, ${\cal O}(n^3)$.

\end{itemize}
Part of our planned future work is to gain some theoretical understanding
of these facts.  Furthermore, we conjecture that a more sophisticated
hybridization of these three heuristics could lead to even better
results.

\section{Acknowledgements}

We thank the anonymous referees for the many suggested improvements of
the paper.  This research was partially funded by Funda\c{c}\~ao para
a Ci\^encia e Tecnologia (FCT) and Program POSI, and by projects ASA
(PTDC/MAT/65481/2006) and CANTE (PTDC/EIA-CCO/101904/2008). Davide
Nabais was funded by a LIACC-FCT scholarship for young undergraduate
researchers.  \bibliographystyle{alpha}

\begin{thebibliography}{EKSW05}

\bibitem[AAA{\etalchar{+}}09]{almeida09:_fado_guitar}
A.~Almeida, M.~Almeida, J.~Alves, N.~Moreira, and R.~Reis.
\newblock {FAdo and GUItar: tools for automata manipulation and visualization}.
\newblock In S.~Maneth, editor, {\em CIAA 2009: 14th International Conference
  on Implementation and Application of Automata}, volume 5642 of {\em LNCS},
  pages 65--74, Sidney, July 2009. Springer.

\bibitem[AH09]{ahn09:_implem_of_state_elimin_using_heuris}
J.-H. Ahn and Y.-S. Han.
\newblock Implementation of state elimination using heuristics.
\newblock In S.~Maneth, editor, {\em CIAA 2009, 14th International Conference
  on Implementation and Application of Automata}, volume 5642 of {\em LNCS},
  pages 178--187, Sidney, July 2009. Springer.

\bibitem[AMR07]{almeida07:_enumer_gener_strin_autom_repres}
M.~Almeida, N.~Moreira, and R.~Reis.
\newblock Enumeration and generation with a string automata representation.
\newblock {\em Theoret. Comput. Sci.}, 387(2):93--102, 2007.
\newblock Special issue "Selected papers of DCFS 2006".

\bibitem[Ard60]{arden60:_delay_logic_and_finit_state_machin}
D.~N. Arden.
\newblock Delayed logic and finite state machines.
\newblock In {\em Theory of Computing Machine Design}, pages 1--35. U. of
  Michigan Press, Ann Arbor, 1960.

\bibitem[BJ63]{brzozowski63:_signal_flow_graph_techn_for}
J.~A. Brzozowski and E.~J.~McCluskey Jr.
\newblock Signal flow graph techniques for sequential circuit state diagrams.
\newblock {\em IEEE Trans. on Electronic Computers}, EC-12(2):67--76, 1963.

\bibitem[DM04]{delgado04:_approx_to_small_regul_expres}
M.~Delgado and J.~Morais.
\newblock Approximation to the smallest regular expression for a given regular
  language.
\newblock In M.~Domaratzki, A.~Okhotin, K.~Salomaa, and S.~Yu, editors, {\em
  CIAA 2004, 9th International Conference on Implementation and Application of
  Automata}, volume 3317 of {\em LNCS}, pages 312--314. Springer, 2004.

\bibitem[EKSW05]{ellul05:_regul_expres}
K.~Ellul, B.~Krawetz, J.~Shallit, and M.~Wang.
\newblock Regular expressions: New results and open problems.
\newblock {\em J. Aut., Lang. and Combin.}, 10(4):407--437, 2005.

\bibitem[FAd10]{fado}
Project FAdo.
\newblock {FAdo}: tools for formal languages manipulation.
\newblock \texttt{http://www.ncc.up.pt/FAdo}, {Access date:1.1.2010}.

\bibitem[GF08]{gulan08:_local_elimin_strat_in_autom_short}
S.~Gulan and H.~Fernau.
\newblock Local elimination-strategies in automata for shorter regular
  expressions.
\newblock In V.~Geffert, J.~Karhum{\"a}ki, A.~Bertoni, B.~Preneel,
  P.~N{\'a}vrat, and M.~Bielikov{\'a}, editors, {\em SOFSEM 2008, Nov{\'y}
  Smokovec, Slovakia, 2008, Volume II - Student Research Forum}, pages 46--57,
  2008.

\bibitem[GH08a]{gruber08:_finit_autom_digrap_connec_and}
H.~Gruber and M.~Holzer.
\newblock Finite automata, digraph connectivity, and regular expression size.
\newblock In L.~Aceto, I.~Damg{\aa}rd, L.~A. Goldberg, M.~MM. Halld{\'o}rsson,
  A.~Ing{\'o}lfsd{\'o}ttir, and I.~Walukiewicz, editors, {\em ICALP 2008, 35th
  International Colloquium on utomata, Languages and Programming, Part II},
  volume 5126 of {\em LNCS}, pages 39--50, Reykjavik, Island, July 2008.
  Springer.

\bibitem[GH08b]{gruber08:_provab_short_regul_expres_from}
H.~Gruber and M.~Holzer.
\newblock Provably shorter regular expressions from deterministic finite
  automata.
\newblock In M.~Ito and M.~Toyama, editors, {\em Proceedings of the 12th
  International Conference Developments in Language Theory}, number 5257 in
  LNCS, pages 383--395, Kyoto, September 2008. Springer.

\bibitem[GHT09]{gruber09:_short_regul_expres_from_finit_autom}
H.~Gruber, M.~Holzer, and M.~Tautschnig.
\newblock Short regular expressions from finite automata: Empirical results.
\newblock In S.~Maneth, editor, {\em CIAA 2009, 14th International Conference
  on Implementation and Application of Automata}, volume 5642 of {\em LNCS},
  pages 188--197, Sidney, July 2009. Springer.

\bibitem[GS07]{gramlich07:_minim_nfas_and_regul_expres}
G.~Gramlich and G.~Schnitger.
\newblock Minimizing nfa's and regular expressions.
\newblock {\em J. Comput. Syst. Sci.}, 73(6):908--923, 2007.

\bibitem[Har69]{harary69:_graph_theor}
F.~Harary.
\newblock {\em Graph Theory}.
\newblock Addison Wesley, 6th edition, 1969.

\bibitem[HMU00]{hopcroft00:_introd_autom_theor_languag_comput}
J.~E. Hopcroft, R.~Motwani, and J.~D. Ullman.
\newblock {\em Introduction to Automata Theory, Languages and Computation}.
\newblock Addison Wesley, 2000.

\bibitem[HW07]{han07:_obtain_short_regul_expres_finit_state_autom}
Y.-S. Han and D.~Wood.
\newblock Obtaining shorter regular expressions from finite-state automata.
\newblock {\em Theoret. Comput. Sci.}, 370:110--120, 2007.

\bibitem[JR93]{jiang93:_minim_nfa}
T.~Jiang and B.~Ravikumar.
\newblock Minimal {NFA} problems are hard.
\newblock {\em SIAM Journal of Computation}, pages 1117--1141, 1993.

\bibitem[Kle56]{kleene56:_repres}
S.~C. Kleene.
\newblock Representation of events in nerve nets and finite automata.
\newblock In C.~E. Shannon and J.~McCarthy, editors, {\em Automata Studies},
  pages 3--41. Princeton University Press, 1956.

\bibitem[Koz94]{kozen94:_kleen}
D.~C. Kozen.
\newblock A completeness theorem for {K}leene algebras and the algebra of
  regular events.
\newblock {\em Infor. and Comput.}, 110(2):366--390, May 1994.

\bibitem[MR05]{moreira05:_inter_manip_regul_objec_fado}
N.~Moreira and R.~Reis.
\newblock Interactive manipulation of regular objects with {FAdo}.
\newblock In {\em Proceedings of 2005 Innovation and Technology in Computer
  Science Education (ITiCSE 2005)}, pages 335--339. ACM, 2005.

\bibitem[MR09]{moreira09:_series_paral_autom_and_short_regul_expres}
N.~Moreira and R.~Reis.
\newblock Series-parallel automata and short regular expressions.
\newblock {\em Fundam. Inform.}, 91(3-4):611--629, 2009.

\bibitem[MY60]{mcnaughton60:_regul_expres_and_state_graph_for_autom}
R.~McNaughton and H.~Yamada.
\newblock Regular expressions and state graphs for automata.
\newblock {\em IEEE Trans. on Electronic Computers}, EC-9(1):39--47, 1960.

\bibitem[Sak05]{sakarovitch05:_languag_expres_small_autom}
J.~Sakarovitch.
\newblock The language, the expression, and the (small) automaton.
\newblock In I.~Litovshy J.~Farré and S.~Schmitz, editors, {\em CIAA 2005,
  10th International Conference on Implementation and Application of Automata},
  volume 3845 of {\em LNCS}, pages 15--30. Springer, 2005.

\bibitem[Sak09]{sakarovitch09:_elemen_of_autom_theor}
J.~Sakarovitch.
\newblock {\em Elements of Automata Theory}.
\newblock CUP, 2009.

\end{thebibliography}
\newcommand{\etalchar}[1]{$^{#1}$}

\end{document}